\documentclass[pra,twocolumn,showpacs,showkeys,secnumarabic,amssymb,amsmath,superscriptaddress,floatfix]{revtex4-1}
\usepackage{amssymb,amsmath}
\usepackage{color}
\usepackage{tikz}
\usepackage{mathtools}
\usetikzlibrary{arrows.meta}
\usepackage{graphics}
\usepackage{graphicx}
\usepackage{enumitem}
\usepackage{bm}
\usepackage{standalone}

\begin{document}
\title{Fragility of quantum correlations and coherence in a multipartite photonic system}

\author{Huan Cao}
\thanks{These authors contributed equally}
\affiliation{CAS Key Laboratory of Quantum Information, University of Science and Technology of China, Hefei, Anhui 230026, China.}
\affiliation{CAS Center For Excellence in Quantum Information and Quantum Physics, University of Science and Technology of China, Hefei, 
Anhui 230026, China.}

\author{Chandrashekar Radhakrishnan}
\thanks{These authors contributed equally}
\affiliation{New York University Shanghai, 1555 Century Ave, Pudong, Shanghai 200122, China.}
\affiliation{NYU-ECNU Institute of Physics at NYU Shanghai, 3663 Zhongshan Road North, Shanghai 200062, China.}
\affiliation{Laboratoire ESIEA Numérique et Société, ESIEA, 9 Rue Vesale, Paris 75005, France.}

\author{Ming Su}
\affiliation{ARC Center for Engineered Quantum systems, School of Mathematics and Physics, University of Queensland, Brisbane, Queensland 4072, Australia.}
\affiliation{NYU-ECNU Institute of Physics at NYU Shanghai, 3663 Zhongshan Road North, Shanghai 200062, China.}

\author{Md. Manirul Ali}
\affiliation{Department of Physics, National Cheng Kung University, Tainan 70101, Taiwan.}

\author{Chao Zhang}
\email{zhc1989@ustc.edu.cn}
\affiliation{CAS Key Laboratory of Quantum Information, University of Science and Technology of China, Hefei, Anhui 230026, China.}
\affiliation{CAS Center For Excellence in Quantum Information and Quantum Physics, University of Science and Technology of China, Hefei, 
Anhui 230026, China.}

\author{Y.F. Huang}
\email{hyf@ustc.edu.cn}
\affiliation{CAS Key Laboratory of Quantum Information, University of Science and Technology of China, Hefei, Anhui 230026, China.}
\affiliation{CAS Center For Excellence in Quantum Information and Quantum Physics, University of Science and Technology of China, Hefei, 
Anhui 230026, China.}

\author{Tim Byrnes}
\email{tim.byrnes@nyu.edu}
\affiliation{New York University Shanghai, 1555 Century Ave, Pudong, Shanghai 200122, China.}
\affiliation{State Key Laboratory of Precision Spectroscopy, School of Physical and Material Sciences,East China Normal University, Shanghai 200062, China.}
\affiliation{NYU-ECNU Institute of Physics at NYU Shanghai, 3663 Zhongshan Road North, Shanghai 200062, China.}
\affiliation{National Institute of Informatics, 2-1-2 Hitotsubashi, Chiyoda-ku, Tokyo 101-8430, Japan.}
\affiliation{Department of Physics, New York University, New York, NY 10003, USA.}

\author{Chuangfeng Li}
\email{cfli@ustc.edu.cn}
\affiliation{CAS Key Laboratory of Quantum Information, University of Science and Technology of China, Hefei, Anhui 230026, China.}
\affiliation{CAS Center For Excellence in Quantum Information and Quantum Physics, University of Science and Technology of China, Hefei, 
Anhui 230026, China.}

\author{Guang-Can Guo}
\affiliation{CAS Key Laboratory of Quantum Information, University of Science and Technology of China, Hefei, Anhui 230026, China.}
\affiliation{CAS Center For Excellence in Quantum Information and Quantum Physics, University of Science and Technology of China, Hefei, 
Anhui 230026, China.}

\begin{abstract}
Certain quantum states are well-known to be particularly fragile in the presence of decoherence, as illustrated by Schrodinger's famous gedanken cat experiment.  It has been better appreciated more recently that quantum states can be characterized in a hierarchy of quantum quantities such  entanglement, quantum correlations, and quantum coherence. It has been conjectured that each of these quantities have various degrees of fragility in the presence of decoherence.  Here we experimentally confirm this conjecture by preparing tripartite photonic states and subjecting them to controlled amounts of dephasing. When the dephasing is applied to all the qubits, we find that the entanglement is the most fragile quantity, followed by the quantum coherence, then mutual information.  This is in agreement with the widely held expectation that multipartite quantum correlations are a highly fragile manifestation of quantumness. We also perform dephasing on one out of the three qubits on star and $ W \bar{W} $ states.  Here the distribution of the correlations and coherence in the state becomes more important in relation to the dephasing location. 
\end{abstract}
 
\maketitle

\section{Introduction}

One of the main challenges in the development of quantum technologies is how to overcome decoherence \cite{steane1998quantum,dowling2003quantum,ladd2010quantum}. Quantum systems tend to couple very easily to their external environment, losing their quantum nature, reducing to a classical state \cite{breuer2002theory,zurek2003decoherence}.  It is however also well-known that the time scale for which a quantum state decoheres is very much a state-dependent process.  For example, superpositions of macroscopically distinct states, such as Schrodinger cat states $ |0 \rangle^{\otimes N}  + |1 \rangle^{\otimes N} $, where $ N $ is the number of qubits, collapse exponentially faster in comparison to a product state of qubits $ (|0 \rangle  + |1 \rangle)^{\otimes N} $.   The fragility (or conversely the robustness) of quantum states have been studied in numerous studies \cite{frowis2018macroscopic,yu2002phonon,shimizu2002stability,janzing2000fragility}. The fragility of quantum states has been discussed in connection to measures of defining the macroscopicity of quantum superpositions \cite{frowis2018macroscopic,frowis2012measures,dur2002effective}.  The fragility of particular quantum states can be considered the flip-side of the enhanced sensitivity of such states, the classic example being NOON states, which are fundamental in the field of quantum metrology \cite{boto2000quantum,dowling2008quantum,giovannetti2011advances}.

Meanwhile, quantum information theory has provided numerous tools in order to better understand the nature of quantum states.  Various quantifiers for strength of Bell correlations \cite{bell1964einstein,einstein1935can}, EPR steering \cite{wiseman2007steering}, entanglement \cite{horodecki2009quantum}, and quantum correlations \cite{ollivier2001quantum,henderson2001classical} have been proposed, each characterizing different aspects of quantum states.  For example, entanglement is strictly defined as any state that is not writeable in separable form, whereas quantum correlations arise when it is impossible to disturb a quantum state with local projective measurements \cite{ollivier2001quantum}.  Recently another quantifier, quantum coherence, has attracted attention as another way of characterizing quantum states \cite{baumgratz2014quantifying}.  Unlike quantum correlations that require at least bipartite systems to exist, quantum coherence can occur on a single system, and is a measure of the degree of superposition \cite{radhakrishnan2016distribution,tan2016unified}. 
These quantifiers form a hierarchical structure, where quantities higher in the hierarchy possess attributes non-zero values of lower quantites \cite{adesso2016measures,ma2019operational}.  For example, a system possessing entanglement necessarily possesses quantum correlations and coherence, but does not necessarily show Bell correlations or steering.  In particular, a unified theory connecting various types of quantum correlations  was proposed by Modi, Vedral, Williamson, and co-workers in Ref. \cite{modi2010unified}. Giorgi and Zambrini extended this approach to include various types of coherence in Ref. \cite{giorgi2018hallmarking}. Various quantum technological tasks rely on different properties of quantum states, hence one of the major aims of quantum information theory is to understand the 
operational capability of these different resources \cite{chitambar2019quantum,winter2016operational,brandao2015reversible,radhakrishnan2017quantum,chitambar2016relating,horodecki2013quantumness,chitambar2016critical}. 
How these resources behave in a dynamical context has been a focus of several works \cite{xu2010experimental,xu2010experimental2,xu2013experimental,bernardes2015experimental}, motivated by the presence of environmental decoherence in quantum technological systems.

In this study, we experimentally show the effect of the different quantum correlations and coherences of a tripartite photonic system under the influence of a one and three qubit dephasing environment.   We measure the six quantities: (1) entanglement, (2) total coherence, (3) global coherence, (4) local coherence, (5) mutual information, and (6) classical correlations and measure their decay dynamics under dephasing.  The fragility of these quantities under dephasing is investigated by measuring the decay rate, which can quantify the fragility of the quantity under question.   We note that investigations on the transient dynamics of entanglement and quantum discord  have been performed in Refs. \cite{bellomo2007non,eberly2007end,lopez2008sudden,yu2009sudden, xu2010experimental,xu2010experimental2,xu2013experimental,bernardes2015experimental,mazzola2010sudden,
werlang2009robustness,maziero2009classical}.  Particularly in Ref. \cite{xu2010experimental,xu2010experimental2,xu2013experimental,bernardes2015experimental} 
an experimental verification of the decay dynamics has been examined. In our work we focus on studying the {\it comparative} dephasing dynamics of different
quantum properties using relative entropy measures.  To observe the decay dynamics of multipartite quantum states,  we generate the $ W \bar{W} $ and star states, which contain correlations and coherences at all levels.  Such states are uniquely suited for examining multiple quantum properties simultaneously.

\begin{figure*}
\includegraphics[width=\linewidth]{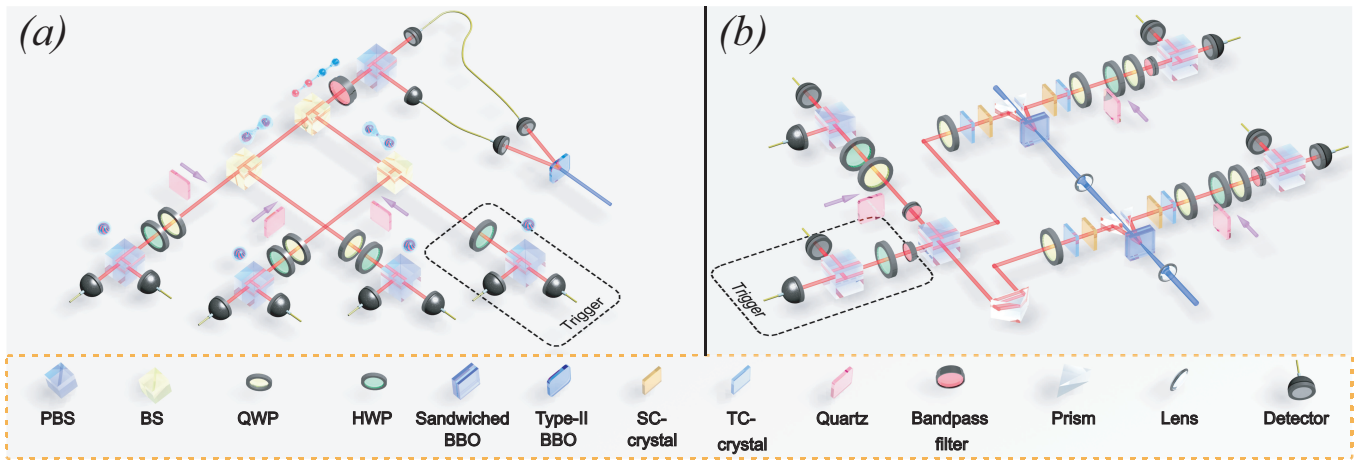}
\caption{Experimental setup for the preparation, dephasing, and measurement of the (a) $W \bar{W}$ state and (b) star state.  In (a), the down-converted 
photons are collected by a single fiber coupler. 
The output coupler before the first polarizing beam splitter (PBS) is mounted on a translational stage to make fine adjustments with the arrival time of the 
photons.  Each beam splitter (BS) consists of one 
$0^\circ $ plate BS and a $45^\circ$ mirror in its reflection path.  The mirror introduces 
a phase shift of $\pi$ between $|H \rangle$ and $|V \rangle$ which is to compensate the phase shift introduced by the BS.  Such a setup makes the reflectivity more polarization independent than a cube BS. Since this requires no phase modulation the setup can be stable over several days. The final triggered photon is detected using a half-wave plate (HWP), a PBS, and a detector.  
Each photon of a $W \bar{W}$ state is analyzed using a polarization measurement system consisting of a quarter wave plate (QWP), HWP, PBS and two fiber coupled single photon detectors. (b) In each arm of the EPR pairs, one TC-crystal is used for temporal compensation and one SC-crystal is used for
spatial compensation, through which the two possible ways of generating photon pairs (first or second crystal in sandwiched BBO) are made 
indistinguishable.  The two extraordinarily down converted photons produced by cascaded sandwich beam 
source is superposed on a PBS. The time of arrival of photons are adjusted with prisms. Further details on the experiment can be seen in the 
Supplementary Material. \label{fig1} }
\end{figure*}

\section{Photonic state generation}
\subsection{$W \bar{W}$ and Star states}

In this study we generate and study the dynamics of two quantum states under dephasing.  The first state is the $W \bar{W}$ state defined as
\begin{align}
|W \bar{W} \rangle &= \frac{1}{\sqrt{2}} \left( |W \rangle + |\bar{W} \rangle \right) \\
|W \rangle  & = \frac{1}{\sqrt{3}} \left( |001 \rangle + |010 \rangle + |100 \rangle \right)  \\
| \bar{W} \rangle &= \frac{1}{\sqrt{3}} \left( |110 \rangle + |101 \rangle + |011 \rangle  \right) 
\end{align}
The $W \bar{W}$ state is an equal superposition of a standard $ W $ state and its spin-flipped version, the $ \bar{W}$ state. This type of state is chosen 
because it has quantum coherence at the single qubit, bipartite and tripartite levels, as well as bipartite and tripartite quantum correlations. Such a state is a good testbed for studying quantum correlations distributed at different levels. The presence of different types of correlations is one of the reasons that $W$ states are robust under local decoherence \cite{dur2000three}.

The second state we investigate is the star state defined as
\begin{align}													
|S \rangle &= \frac{1}{2} \left(|000 \rangle + |100 \rangle + |101 \rangle + |111 \rangle \right).   
\label{starstatedef}                 
\end{align}
Like the $W \bar{W}$ state, the star state also has coherence and correlations distributed at all possible levels.  However, the correlations are present in an asymmetric way for a star state, in contrast to the $W \bar{W}$ state which is symmetric for all qubits. The entanglement structure for the star state takes a form $ A \Leftrightarrow C \Leftrightarrow B $, where we have labeled the three qubits as $ ABC $ in (\ref{starstatedef}) from left to right.   For example, if qubits $ A $ or $ B $ are traced out, entanglement is present in the remaining qubits.  However, if qubit $ C $ is traced out, the remaining qubits are left in a separable state.  We thus call qubit $C$ is the central qubit, and qubits $ A $ and $ B $ the peripheral qubits. The star state is a very simple example of a graph state \cite{plesch2003entangled} which in multipartite cases are 
useful for quantum error correction \cite{anders2006fast}. More details on the distribution of correlations and coherence in the $W \bar{W}$ and star states are given in the Supplementary Material.  

\subsection{Experimental Preparation}

To experimentally realize the above states, polarization encoded photonic qubits
are used, where the horizontal (H) and vertical (V) polarizations are encoded as the two levels $|0 \rangle$ and $|1 \rangle$ respectively.  The detailed procedure of preparing these quantum states is shown in Fig. \ref{fig1}. In our experiment we investigate the dynamics of various correlations and coherence in a tripartite quantum system which is under the influence of an external phase damping environment, realized by passing the photonic states through birefringent quartz crystals of different thicknesses. We perform two types of dephasing, where all three photons are dephased by a crystal of the same thickness, and another where only one of the photons is dephased.  The dephasing on only one of the photons allows for a partial dephasing of the system, where some quantum property is retained even after complete dephasing.

The experimental set up to prepare the $W \bar{W}$ is shown in Fig. 1(a).  Two pairs of down converted photons are simultaneously generated through  a 
higher order emission of spontaneous down conversion (SPDC) process.  These four photons are collected by a single mode fiber and then fed into a 
polarizing beam splitter (PBS) where they overlap and become indistinguishable in the spatial mode.  The spectral selection is realized by inserting a 3nm 
interference filter after the PBS.  The four photons are separated by three non-polarizing beam splitters (BS).  The post selected four-fold coincidence 
count certifies the generation of four photon Dicke state with two excitations  
$|D_{4}^{2} \rangle = (|0011 \rangle + |0101 \rangle +  |1001 \rangle + |0110 \rangle  + |1010 \rangle + |1100 \rangle) /\sqrt{6}$.  
The $W \bar{W}$ state is generated from the Dicke state by projecting one of the qubits into the 
$(|0 \rangle + |1 \rangle)/\sqrt{2}$ basis.

The star state generation scheme is shown in Fig. 1(b).  Two non-maximally entangled bipartite states
$|\psi \rangle = \cos \theta |01 \rangle + \sin \theta |10 \rangle$ with the ratio
$\cos^{2} \theta$:$\sin^{2} \theta$ $=6.8554$, are required to prepare the star state. These polarization entangled states are generated using a sandwiched geometry beam-like type II BBO entanglement resource.  Such an entanglement resource was first devised by Zhang and co-workers in Ref. \cite{zhang2015experimental} and was later used 
in Ref. \cite{wang2016experimental} to realize ten photon entanglement. Applying single qubit unitary operators on each qubit, the state $|\psi \rangle$ is transformed to $(|00 \rangle + |10 \rangle + |11 \rangle)/\sqrt{3}$.  The transformed states are fed into the PBS to overlap them and the Hong-Ou-Mandel interference visibility is enhanced using a 2nm band pass filter.  The second one of the four qubit quantum states generated through this process is projected in the 
$(|0 \rangle + |1 \rangle)/\sqrt{2}$ basis. By exchanging the qubits $3$ and $4$ in the resulting quantum state, the star states are obtained.

\begin{table*}
\begin{center}
\begin{tabular}{|c|c|c|c|}
\hline
\text{Quantity} & \text{Reference state} & Example reference state &  \text{Definition} \\
\hline
\text{Entanglement} & \text{Separable state \; $\mathcal{S}$} & $ \sum_j p_j \rho^A_j \otimes \rho^B_j \otimes \rho^C_j $ &  $ E = \displaystyle{\min_{\sigma \in \mathcal{S}}} \;  S(\rho \| \sigma)$\\
 \hline
 \text{Total Coherence} & \text{Incoherent state \; $\mathcal{I}$} & $\rho_d = \sum_{j} \langle j | \rho | j \rangle  |j \rangle \langle j | $ &  
$C = \displaystyle{\min_{\sigma \in \mathcal{I}}}  \;  S(\rho \| \sigma) $  \\
\hline
\text{Local Coherence} &  Incoherent states  $ \mathcal{I} $   & $\pi_d(\rho) = \rho^A_d \otimes \rho^B_d \otimes \rho^C_d $ & 
$ C_{L} = \displaystyle{\min_{ \pi(\rho) \in \mathcal{I}}} S(\pi(\rho) \| \sigma ) $ \\
\hline
$\underset{\text{(Total correlations)}}{\text{Mutual Information}}$ & \text{Product state \; $\mathcal{P}$} & $ \pi(\rho) = \rho^A \otimes \rho^B \otimes \rho^C $ & 
$T= \displaystyle{\min_{\sigma \in \mathcal{P}}} \;  S(\rho \| \sigma) $ \\ 
\hline
\text{Classical correlation} &\text{Product state \; $\mathcal{P}$}  & $ \pi(\rho_d) $ &  
$ K = \displaystyle{\min_{\sigma \in \mathcal{P}}} S(\rho_d \| \sigma ) $ \\
\hline
\text{Hookup} & \text{Incoherent product states  $\bar{\mathcal{I}}$} & $ \pi_d(\rho) = \pi(\rho_d) $  & 
$M = \displaystyle{\min_{\sigma \in \bar{\mathcal{I}}}}  \;  S(\rho \| \sigma) $ \\
\hline 
\end{tabular}
\end{center}
\caption{List of properties of a quantum state $\rho$ and their measurement procedure. \label{table1} }
\end{table*}

\section{Measures of correlations and coherence}

We measure the correlations and coherence using the unified distance-based approach of Ref. \cite{modi2010unified}.  The basic idea of any distance-based approach to quantify a quantum observable is as follows.  First the set of all states do not have the relevant quantity is defined, and are called the reference states.  For example, for entanglement, the reference states are the set of all separable states.  Then to quantify the quantum property, one uses a suitable distance measure to find the distance to the closest reference state by minimization.  In our case, we choose the distance measure to be relative entropy
\begin{equation}
S(\rho \| \sigma) = \hbox{Tr} (\rho \ln \rho - \rho \ln \sigma),
\label{REdensitymatrices} 
\end{equation}
which is a popular choice due to its simplicity of computation and well-known properties \cite{vedral2002role}. The six quantities that we calculate are defined as below and summarized in Table \ref{table1}. 
 
{\it Entanglement:} The entanglement is quantified as the minimum distance to the set of all separable states \cite{vedral1997quantifying,vedral1998entanglement}.  We perform a minimization procedure to separable states taking the form  $ \sum_j p_j \rho^A_j \otimes \rho^B_j \otimes \rho^C_j $, where $ p_j $ is a probability and $  \rho^{A,B,C}_j $ are density matrices on subsystems $ A, B, C $.  

{\it Coherence:} 
The total quantum coherence \cite{baumgratz2014quantifying} is defined as the distance to the closest incoherent state, which take the form  $  \sum_{j}p_j  |j \rangle \langle j | $, where $ |j \rangle $ are in the basis $ \{ |0 \rangle, |1 \rangle \} $ for $ A, B, C $. It has been shown that for the relative entropy, the closest incoherent state to a state $ \rho $ takes coefficients $ p_j = \langle j | \rho | j \rangle $, hence the minimization does not need to be explicitly performed \cite{baumgratz2014quantifying} and 
\begin{equation}
C(\rho) =  \min_{\sigma \in \mathcal{I}} S(\rho \| \sigma) = S(\rho \| \rho_d ) =  S(\rho_{d}) - S(\rho).  
\label{totalcoherence}
\end{equation}
Here we defined $ \rho_d $ as the matrix $ \rho $ with all off-diagonal terms set to zero in the basis $ | j \rangle $.  

{\it Local and global coherence:} 
Quantum coherence can originate from both coherence which is localized on subsystems, or due to coherence due to a collective property of the whole system \cite{radhakrishnan2016distribution,tan2016unified}.  The former is called local coherence and is found by first breaking all the correlations between the subsystems.  In a similar way to total coherence, the closest incoherent state is found by taking the diagonal form 
\begin{align}
C_{L}(\rho) & = \min_{\sigma \in \mathcal{I}} S(\pi(\rho) \| \sigma )  =  S(\pi(\rho) \| \pi_d (\rho) ) ,
\end{align}
where $ \pi_d (\rho) $ is the matrix $ \pi(\rho) $ but with all off-diagonal elements set to zero in the basis $ | j \rangle $.  The coherence attributed to the collective nature of the system is called the global coherence and is defined as the difference of the total and local coherence
\begin{align}
 C_{G} (\rho)  & = C(\rho) - C_{L}(\rho) .  
\end{align}

{\it Mutual Information:} 
Mutual information measures the total amount of correlations, including both quantum and classical parts  \cite{modi2010unified}.  The set of uncorrelated states takes the form of a product state 
$ \sigma^A \otimes \sigma^B \otimes \sigma^C $.  It has been shown in Ref. \cite{modi2010unified} that for relative entropy the closest product state is the product state $ \pi (\rho) =  \rho^A \otimes \rho^B \otimes \rho^C $ consisting of the reduced density matrices on each subsystem $ \rho^{A,B,C} $.  Hence we can write 
\begin{equation}
T(\rho) =  \min_{\sigma \in \mathcal{P}} S( \rho \| \sigma )  =  S( \rho \| \pi(\rho)) \equiv S(\pi(\rho)) - S(\rho) .
\label{mutualinformation}
\end{equation}
The total correlations as measured by the mutual information $T$ and the total quantum coherence $C$ are not completely independent quantities. Hence there is a common region of quantumness in a system which is measured by both these quantities.  This region of overlap is the amount of global coherence in the system which arises due to quantum correlations between the qubits.

{\it Classical correlations:}
For local coherence, first the correlations between the subsystems are broken, then the remaining coherence is measured.  The reverse ordering can equally be performed, where first the coherence is removed from the system, then the remaining correlations are measured.  The state with no coherence is $ \rho_d $, which can only contain classical correlations because it is a diagonal density matrix \cite{giorgi2018hallmarking}. In the same way as mutual information, the closest uncorrelated state is its corresponding product state, 
\begin{equation}
K(\rho) = \min_{\sigma \in \mathcal{P}} S(\rho_d \| \sigma ) = S(\rho_{d} \| \pi(\rho_{d})).
\label{classicalcorrelations}
\end{equation}

{\it Hookup:}
The reference state for total coherence $ C $ is $ \rho_d $, which is a state which has no coherence, but potentially classical correlations.  Meanwhile, the reference state for the mutual information $ T $ is $ \pi (\rho) $, which has no correlations but potentially coherence.  One can define a quantity with a reference state that has no correlations and no coherence.  This was called the ``hookup'' in Ref. \cite{giorgi2018hallmarking} and can be evaluated to be  
\begin{equation}
M(\rho) = C(\rho) + K(\rho) = T(\rho) + C_{L} (\rho).
\label{hookup}
\end{equation}
A detailed overview on the various correlations and coherence is given in the Supplementary Material.

\begin{figure*}
\includegraphics[scale=0.060]{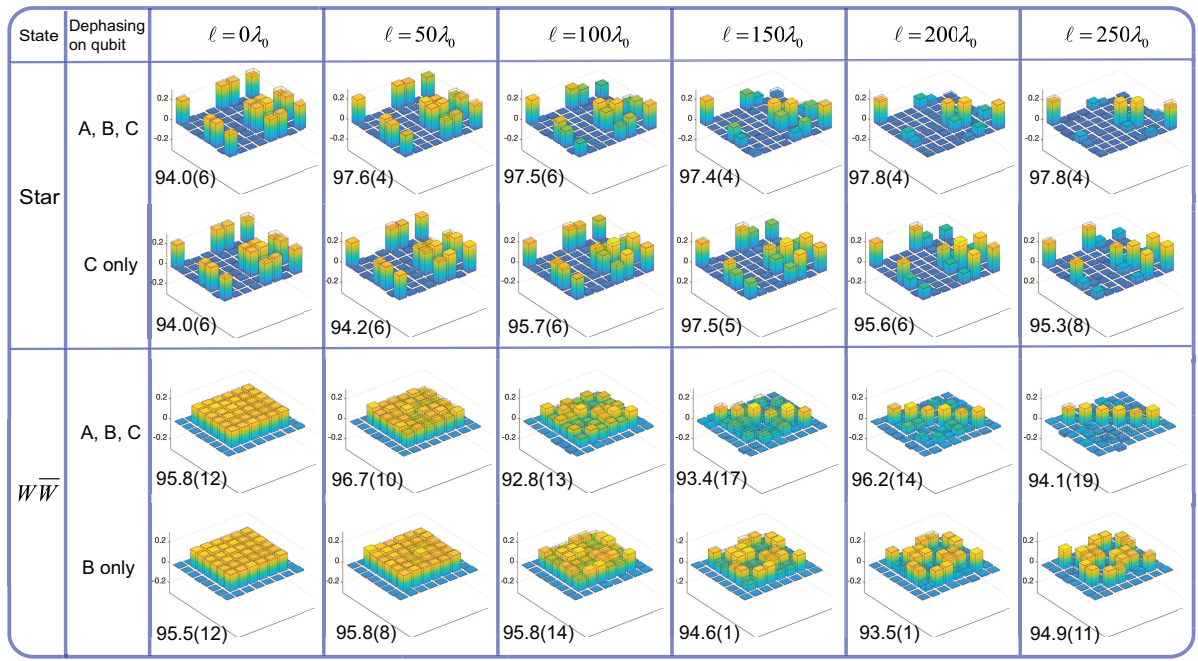}
\caption{Tomographic reconstruction of the density matrices of the star and $W \bar{W}$ states for various thicknesses of quartz plates.  
The amount of dephasing is controlled by the quartz plate thickness $ \ell $. Only the real part of the density matrix elements are shown, 
and the imaginary parts are consistent with zero for all thicknesses (see Supplementary Material).  
The theoretical density matrix for each dephasing time is shown as a transparent histogram, and the fidelities are marked as a percentage, along with the error estimate. 
Dephasing rates of $ \Gamma  = 2.21 \times 10^{-5} \lambda_{0}^{-2} $ for the $ W \bar{W} $ and 
$ \Gamma = 2.06 \times 10^{-5} \lambda_{0}^{-2} $ for star states are used, with $\lambda_{0} = 780$ nm.  
\label{tomography}  }
\end{figure*}

\section{Density matrix evolution under dephasing}

\subsection{Tomography reconstruction of states}

Fig. \ref{tomography} shows the tomographic reconstructions of the star and $ W \bar{W} $ states with various amounts of dephasing. For the case that the dephasing is applied to all the photons, the density matrix approaches its diagonal form as expected for larger values of $ \ell $, the thickness of the quartz plate. The case where dephasing is only applied to one of the photons, off-diagonal terms remain since the state is only partially dephased.  This is due to the nature of the star and $ W \bar{W} $ states that are used which contain other types of coherence other than completely tripartite coherence (such as in a GHZ state).  
The tomographically reconstructed density matrix is compared to the theoretically calculated density matrix according to a dephasing channel for each qubit defined as 
\begin{align}
\rho \rightarrow (1- p(\ell)) \rho + p(\ell) \sigma_z \rho \sigma_z ,
\label{dephasingch}
\end{align}
where $p(\ell) = [1-\exp(-\Gamma \ell^{2})]/2$ (see Supplementary Material). We obtain fidelities of the state with dephasing better than $ 93 \% $ for all dephasing values.

\subsection{Decay of correlations and coherence with dephasing on all qubits}

\begin{figure*}
\includegraphics[width=\linewidth]{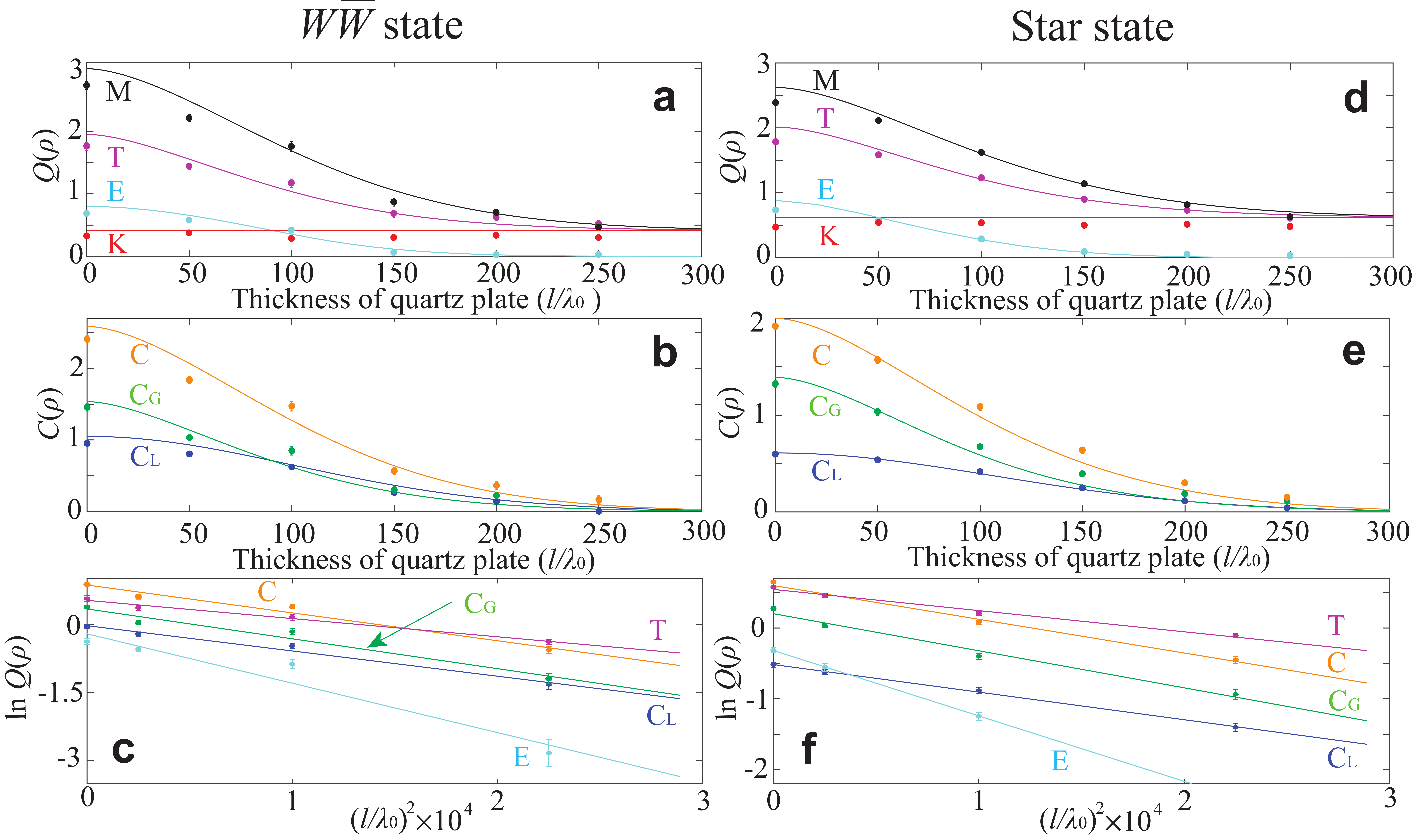}
\caption{Decay of quantum properties for the (a)(b)(c) $W \bar{W}$ and (d)(e)(f) star state under three qubit dephasing.  The various quantum properties are mutual information $ T $, total coherence $ C $ global coherence $ C_G $, local coherence $ C_L $, Entanglement $ E $  and the classical correlations $ K $.  In (a)(b)(d)(e), the exponential decay of these properties is shown as a function of the thickness of the quartz plate $\ell$
(units of $\lambda_{0} =780$nm).  Theoretical predictions are shown with the solid lines.  
In (c)(f), we replot the same curves on a semilog plot with the $x$-axis representing the square of the thickness of the quartz plate and the physical properties along the $y$-axis.  The slope of the linear fit gives the decay rate of the quantum property.  In all figures, the experimental data is denoted by points and the error bar is given by through a simulation of the photon statistics. In (a)(b)(d)(e), solid lines are the theoretical predictions, while in (c)(f) the solid lines are fits to the experimental data.  
Fitted values of the decay rates (in units of $ 10^{-5} \lambda_{0}^{-2} $) are (c) $ \Gamma(E) = 10.9, \Gamma(C_G) = 6.6, \Gamma(C) = 6.1, \Gamma(C_L) = 5.6, \Gamma(T) = 4.0 $; (f) $ \Gamma(E) = 9.2, \Gamma(C_G) = 5.2, \Gamma(C) = 4.8, \Gamma(C_L) = 3.9, \Gamma(T) = 3.0 $.  \label{fig3}}
\end{figure*}

Using the tomographically reconstructed density matrices we calculate the various quantities summarized in Table \ref{table1}.  First we discuss the 
dephasing dynamics of the correlations in a $W \bar{W}$ state, as shown in Fig. \ref{fig3}(a) and (b).  We observe that all quantities decay to zero 
for large dephasing, except the mutual information $T$ and classical correlations $K$, which saturate to finite values.  This is due to the dephasing 
removing all coherence from the system, such that the state
\begin{align}
\rho_d  &=& \frac{1}{6} \Big( |001\rangle \langle 001| + |010\rangle \langle 010| + |100\rangle \langle 100|   \nonumber \\
             &  & +  |110\rangle \langle 110| +  |101\rangle \langle 101|+  |011\rangle \langle 011| \Big)
\end{align}
is progressively approached.  This is a classically correlated state and hence the mutual information only contains classical correlations $ T = K $ as 
observed, and all other quantum properties decay to zero.     
In Fig.  \ref{fig3}(b), we see that the global coherence starts at a larger value than the local coherence, but the global coherence decays faster 
than the local coherence.  This is an indication of the greater robustness of the local coherence in the presence of dephasing than global 
coherence.

To examine this point in more detail, we plot the decay rates for the various quantities in Fig.  \ref{fig3}(c).  Due to the Gaussian nature of the dephasing 
channel (\ref{dephasingch}), we expect the quantum properties to also approximately follow a Gaussian form $ \propto \exp( - \Gamma \ell^{2} ) $, 
hence the decay rate 
is the negative gradient on a semilog plot with $ \ell^2 $.   Of all the 
quantum properties the fastest decay is for entanglement.  The next fastest decay rate is displayed by global quantum coherence, followed by the total 
coherence.   The very slow decay of mutual information is because it is composed of both 
quantum correlations and classical correlations.  While quantum correlations decay due to the environment, the classical correlations
remains unchanged, since the dephasing acts in the classical basis $ |0 \rangle, |1 \rangle $. Likewise, local coherence can be seen to decay more slowly than the total coherence.  These results generally show that the quantities that are related to collective effects, such as entanglement and global coherence tend to decay at a faster than classical or local quantities.  

The star state generally shows similar behavior, as seen in Fig. \ref{fig3}(d) and (e).  Here again the mutual information and classical correlations saturate towards a non-zero value, according to the classical correlations in the state
\begin{align}
\rho_d  = \frac{1}{4} \left(  |000\rangle \langle 000| + |100\rangle \langle 100| + |101\rangle \langle101| + |111\rangle \langle 111| \right) . 
\end{align}
All other quantities decay to zero, in a similar way to the $ W \bar{W} $ state.  The total coherence is less in the star state due to the smaller number of terms in the superposition.  Nevertheless, as seen by evaluating the decay rates in Fig. \ref{fig3}(f), the entanglement shows the greatest rate of decrease, followed by the global and total coherences.  The mutual information and local coherences decay with the slowest rates, similar to the $ W \bar{W} $ state.  Thus despite the rather different structure of the states, a consistent picture emerges once the decay rates are examined.

\begin{figure*}
\includegraphics[width=\linewidth]{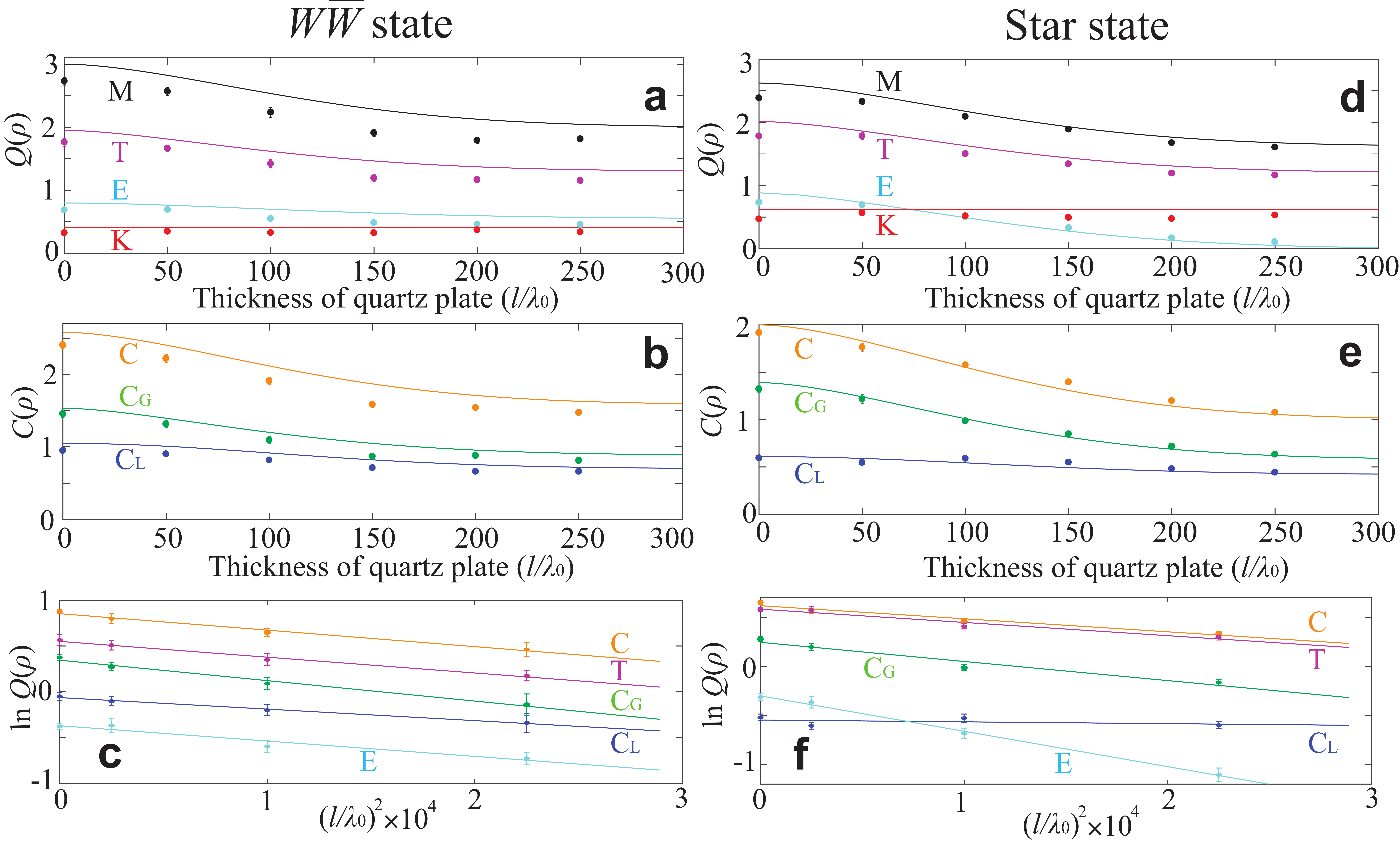}
\caption{Decay of quantum properties of the  (a),(b),(c) $W \bar{W}$ and (d),(e),(f) star state with dephasing of one qubit. For the $W \bar{W}$ state, qubit $ B $ is dephased, while for the star state qubit $ C $ is dephased.  The labeling is the same as for Fig. \ref{fig3}.  Points are experimental data, and the lines are theoretical predictions in (a)(b)(d)(e).  In (c)(f) the lines are fits to the data. Fitted values of the decay rates  (in units of $ 10^{-5} \lambda_{0}^{-2} $) are (c) $\Gamma(C_G) = 2.2,  \Gamma(C) = 1.8,  \Gamma(T) = 1.7,  \Gamma(E) = 1.7, \Gamma(C_L) = 1.3  $; (f) $ \Gamma(E) = 3.6, \Gamma(C_G) = 2.0, \Gamma(C) = 1.3,  \Gamma(T) = 1.3, \Gamma(C_L) = 0.2 $. \label{fig4}  }
\end{figure*}

\subsection{Decay of correlations and coherence with one qubit dephasing}

One way of understanding the faster decay of the collective quantities such as entanglement and global coherence is that they are exposed to the dephasing effects from multiple qubits.  This is in contrast to quantities that are localized on each qubit, such as local coherence, which can only affect one qubit at a time.  In this picture, if the dephasing is only applied to one qubit, then we might expect that the rates for all quantities will be more similar.   To test this hypothesis, we also perform dephasing on one qubit and investigate its effect on the various quantities as before. 

The decay of various quantities for the $W \bar{W}$ due to dephasing is shown in Fig. \ref{fig4}(a) and (b). 
Due to the symmetric nature of the state, dephasing any one of the three qubits leads to the same result, hence in our case qubit $B$ is dephased. 
In this case all quantities saturate to a non-zero value, which is characteristic of the $W \bar{W}$ state.  As is well-known, dephasing of a $W$ state only partially removes the entanglement from the system, and the remaining qubits are partially entangled.  This means that both quantum correlations and coherence are preserved in the system. Due to the quantum correlations that are preserved in this case, we observe that the amount of correlations and coherence are always larger than the amount of classical correlations, in contrast to the three qubit dephasing case.  

The entanglement structure of the state plays a more important role in the case of star states, as seen in Fig. \ref{fig4}(d) and (e).  For the star state we show the effects of dephasing on the central qubit $ C $.  In this case we observe the entanglement decaying to zero for large dephasing, as expected from the discussion surrounding Eq. (\ref{starstatedef}).  For dephasing on a peripheral qubit, we find that the entanglement does not decay to zero, in a similar way to the $ W \bar{W} $ state  (see the Supplementary Material).  Other quantities saturate to non-zero values, with the steady state value of the global coherence being higher than the amount of classical correlations in the system.  This is in 
contrast to the entanglement and local coherence in which the steady state value is lower than the classical correlations. We note that compared to the other
quantities the local coherence exhibits very minimal evolution due to dephasing.

Fig. \ref{fig4}(c) and (f) shows a comparison of the decay rates of the various quantities, which appears as the negative gradient on the semilog plot.  We find that the ordering of the decay rates do not occur in a consistent order as before. For the $ W \bar{W} $ state, we find that all quantities generally decay with a similar rate, with the global coherence giving the largest value. On the other hand, for the star state, we clearly see that the entanglement decays at the fastest rate, in a similar way to the three qubit dephasing case.  We attribute this to the different structure of entanglement that is present in the two states.  For the $ W \bar{W} $ state, all the qubits can be considered ``peripheral'' qubits, since the dephasing only causes partial loss of entanglement.  In the case of dephasing the central qubit of the star state, the destruction of entanglement is very effective, since it is a central qubit for the entanglement.  
Thus in this case we observe that the structure of the quantum correlations greatly affect the fragility of the state.

\section{Summary and Conclusions}

The effects of dephasing on quantum correlations and coherence was experimentally studied on photonic $W \bar{W}$ and star states, with one and three qubit dephasing.  Such states have coherence and correlations of all types in a tripartite system.  Using a Gaussian dephasing model, we are able to extract the effective decay rates for each state and each type of dephasing, as shown in Figs.  \ref{fig3}(c) and (f) and  \ref{fig4}(c) and (f).  In the case that dephasing is applied on all the qubits, a consistent picture emerges, despite the different nature of the states. Here we find that 
\begin{align}
\Gamma(E) > \Gamma(C_G) > \Gamma(C) > \Gamma(C_L) > \Gamma(T) > \Gamma(K) ,  
\label{robustnessordering}
\end{align}
i.e. the dephasing rates occur in the order of entanglement, global coherence, total coherence, local coherence, mutual information, and classical correlations.  
We thus see a clear hierarchy in the decay rate of the various quantum properties, where the collective quantities decay at a faster rate than local and classical quantities.  This can be understood as the result of collective quantities being affected by all the channels of dephasing, but local quantities only being affected by its local dephaser.  In this way we verify the conjecture that collective quantities are more fragile than the local quantities, when local decoherence is applied on the whole system.  For the case that only one qubit is dephased, the rates of decay depend more on the structure of the quantum state. In the case of dephasing the central qubit of a star state, we again recover the entanglement is the most fragile quantity.  However, in the case of dephasing a peripheral qubit where entanglement can be retained in the strong dephasing limit, the rate of decay is much lower. Similar results were obtained in different models of dephasing theoretically \cite{radhakrishnan2019time}.

The hierarchy between the different measurable quantities in (\ref{robustnessordering}) can be viewed as a 
robustness hierarchy in terms of quantum properties as follows
\begin{equation}
\hbox{NLQC} \prec \hbox{TQC} \prec \hbox{LS} .  
\end{equation}
In the above equation NLQC, TQC and LS stand for nonlocal quantum correlation, total quantum correlation and local superposition respectively
and the notation $A \prec B$ denotes that $A$ decays faster than $B$.  The NLQC is unique to quantum systems and 
TQC (both nonlocal and local quantum correlations) are inter-qubit correlations distributed between the qubits.
LS is the superposition between the levels of a qubit and hence is a intra-qubit property which is localized within a qubit.  
Hence we find that the inter-qubit quantum properties which are spread out between the qubits are more likely to decay much faster 
when compared to the intra-qubit quantum properties which are relatively more robust.  This suggest that in quantum information theoretic tasks it would be advantageous to use intra-qubit quantum properties as resources as they can be preserved over longer time intervals.  By converting between local coherence to global coherence only when it is needed \cite{wu2018experimental}, this could be used as a strategy for preserving coherence to longer times.


We note that in our approach the classical correlations are constant throughout the entire process of evolution. This is contrast to the theoretical results observed in Refs.  \cite{maziero2009classical,maziero2010system} and were subsequently experimentally examined \cite{xu2010experimental2}. The difference here originates from the different notions of classicality as defined by quantum discord and quantum coherence.  In quantum discord, a state is classical correlated if there exists a local measurement and a conditioned measurement, in any basis, which does not disturb the quantum state \cite{ollivier2001quantum,radhakrishnan2019quantum}. It is therefore a quantity that is invariant under local basis transformations.  In contrast, coherence is a basis dependent quantity \cite{radhakrishnan2019basis}.  The classical nature of the state is with respect to a particular basis choice, in our case the $ |0\rangle, |1 \rangle $ basis.  Here our notion of classical correlations is in this fixed basis choice, and the dephasing removes coherence in this basis.  This means that the classical correlations are always unchanged under this evolution.  In the case of Refs. \cite{maziero2009classical,maziero2010system,xu2010experimental2}, classical correlations can be dynamic because of the local basis optimization that is performed in evaluating the discord. In our view, these results are not inconsistent, but arise from different notions of classicality.  In our approach, there is a preferred classical basis $ |0\rangle, |1 \rangle $, which is natural to consider since this is the basis that dephasing occurs in the system. 

Another observation that can be made from Fig. \ref{fig3} and \ref{fig4} is that the amount of total quantum coherence is always higher than the entanglement present in the system.  This is 
because the coherence originates due to nonlocal quantum correlations, local quantum correlations and
local superpositions, whereas the entanglement arises only due to the nonlocal quantum correlations.  
This enables us to verify the theorem $E(\rho) \leq C(\rho)$ in Ref. \cite{streltsov2015measuring} in a 
dynamical scenario, when they are both measured using the same contractive distance. 
This relationship between entanglement and coherence was proved in Ref. \cite{streltsov2015measuring} 
 under the condition that both these quantities are measured using the same contractive distance.  In our work we also use the same contractive distance (relative entropy) and  also verify that the relation holds under dephasing dynamics as well. 

Our work demonstrates that various quantum information quantities can be used to effectively characterize quantum systems. These can be extended to larger quantum systems, where more dramatic phase transition phenomena can be observed \cite{radhakrishnan2017quantum}.  Adding decoherence and observing the dynamics can be a direct quantifier for the fragility of various quantities.  Since particular quantities are more relevant for a given quantum information task, this general method may find also practical uses in the context of applications to quantum technology.

\section*{Acknowledgements}
This work is supported by the Shanghai Research Challenge Fund; New York University Global Seed Grants for Collaborative Research; National Natural Science Foundation 
of China (61571301,D1210036A); the NSFC Research Fund for International Young Scientists (11650110425,11850410426); NYU-ECNU Institute of Physics at NYU Shanghai; 
the Science and Technology Commission of Shanghai Municipality (17ZR1443600); the China Science and Technology Exchange Center (NGA-16-001); and the NSFC-RFBR 
Collaborative grant (81811530112).



\end{document}